\documentclass[12pt,preprint]{aastex}
\begin{document}
\title{ORBITAL PERIODICITIES IN THE HARD EMISSION FROM X-RAY 
TRANSIENT - BLACK HOLE CANDIDATES BETWEEN OUTBURSTS}

\author{M.I. Kudryavtsev} 

\affil{Space Research Institute, Russian Academy of Science, Profsoyuznaya st. 84/32, Moscow 117810, Russia}

\author{S.I. Svertilov and V.V. Bogomolov }

\affil{Skobeltsyn Institute of Nuclear Physics, Moscow State University, Vorob'evy Gory, Moscow 119899, Russia}

\begin{abstract}
During long-term observations of the Galactic Centre region in hard X-rays (10-300 keV) in space 
experiments made on board Prognoz-9 satellite and "Mir" orbital station (GRIF experiment) some periodic 
sources were revealed. They include periodicities of hour and day range of period: 152 h, 98 h, 82 h 
(4U1700-37), 69 h, 62 h, 13.3 h, 9.36 h, 8.03h, 8.15 h (Cen X-4), 4.38 h, 4.35 h (4U1755-33) and 3.45 h. For 
all the observed periodic processes the mean phase profiles (light curves) in the different energy ranges were 
obtained. The mean phase profiles of most of the observed periodic sources differ from both sine and purely 
eclipse form. Among six sources with day ranges of period at least three were identified with X-ray Novas - 
black hole candidates: 152 h (H1705-25, Nov.Oph., 1977), 62 h (GRO J 1655-40, Nov. Sco., 1994), 13.3 h 
(4U1543-47). The results of the GRO J 1655-40 and 4U1543-47 observations in the Prognoz-9 and GRIF 
"Mir" experiments were compared with the observational data obtained in space observatories CGRO, 
RXTE, BeppoSax. It should be noted that the orbital periodicity is revealed in the hard emission of the X-ray 
transient - black hole candidates GRO J 1655-40 and 4U1543-47 even in the epochs between outbursts. 
Periodic processes with day ranges of periods can be typical for the transient source like X-ray Nova - black-
hole candidate. Such periodicity is orbital but its origin is not connected with eclipses of the compact 
companion of a binary system. Periodic processes in soft and hard X-rays can be associated with orbital 
motion of binary system companions, but the physical mechanisms and the regions of generation of soft and 
hard X-rays can be quite different.
\end{abstract}


\section{Introduction}
	The enigma of black holes is one of the most intriguing problems of modern astrophysics. It 
is well known that black holes (BH) as well as neutron stars (NS) are associated with hard X-ray 
sources, which are characterized by great energy release as the result of accretion. The temporal 
behavior of such sources was always a problem of great interest for high-energy astrophysicists. 
Modern observations are usually made in a wide range of wavelengths from radio waves to gamma 
rays. While optical observations are in some aspects more accurate, the observation of X-ray 
emitting objects in hard radiation allows us to study the processes in the region of great energy 
release. Moreover, the information on temporal features is necessary for the exact identification of a 
hard source. 

	Various instabilities in the accretion process lead to the flaring activity of many hard X-
ray sources. The properties of matter and fields surrounding a compact object determine quasi-
periodic variations, while the rotation of binary system components produce orbital and super-
orbital periodic changes in X-ray source luminosity. These factors are significantly different in 
dependence on characteristic time. The short-term variability is caused, particularly by quasi-
periodic oscillations (QPO), the properties of which in some cases are different for two classes of 
objects - X-ray binaries with neutron star (NSXB) and with Black Hole (BHXB). Only NS can 
produce QPO with frequency up to 1 kHz \citep{sun00}. Long-term variability of most of BHXB-s is caused 
by their flaring activity. It is known that most of soft X-ray transients (X-ray nova) manifest 
recurrent flares. The typical duration of an X-ray flare is about 50-100 days and the time interval 
between flares varies around 1.5 years. However, a number of X-ray nova flared only once during all 
the period of observations. It is not yet clear, whether there are any soft X-ray transient objects with 
no recurrent flares. The shape of X-ray flares strongly varies from source to source and even from 
one flare to another in the same source. Most of them have standard shape with a "fast rise and an 
exponential decay" but other forms are more complicated with several maximums. An outstanding 
example is the 1994-1997 activity of BHXB GRO J1655-40, which demonstrated a large series of 
quasi-periodic flares, separated from one another by several months \citep{zha94,tav96,zha97,hyn98}.
 
	Variations with minute's, hour's and day's characteristic time scales can occur when X-rays 
are produced or absorbed in some large structures surrounding the compact object as well as in the 
compact object itself. Accretion instabilities or the motion of binary system components may cause 
them. Sometimes they can appear as regular variations of spectra or dips observed at the determined 
orbital phases. The study of such medium-scale temporal phenomena is of great importance. 
Monitoring observations are the traditional way of studying temporal phenomena in hard X-rays and 
soft gamma rays such as periodic processes, outbursts and transients. Periodic variations of the hard 
emission from some galactic sources were discovered long ago. The first observations were made 
on the UHURU space observatory \citep{gia71,for78}, and then studied in 
details during several missions (see for example, \citep{cor86,nag89,pre87} and others). 
Nevertheless, to the present time not so many objects manifesting as periodic 
sources specifically in the hard range of electromagnetic spectrum are known. This is caused by 
some methodical problems, in particular, the necessity of long-term continuous observations of each 
source, which is rather difficult even in the case of monitoring observations particularly for long-
period processes of the orbital and accretion disk precession type. 
Extensive data on periodicity in hard emissions from galactic sources were obtained during 
the BATSE experiment onboard the Compton Orbital observatory (CGRO) \citep{rob97}. 
A study of temporal phenomena in hard X-rays and soft gamma rays was also made in the 
OSSE CGRO experiment \citep{joh93}. More or less regular observation of the Galactic 
Centre region (in the limits of $\sim10^o$x$10^o$) were carried out for over 5 years by means of the 
"SIGMA" telescope on-board the "GRANAT" observatory. This experiment allowed to detect new 
sources at energies of about 50-100 keV and to give the notion about the temporal properties of 
these and previously known objects \citep{gol94}. Currently monitoring observations of 
high-energy properties of X-ray sources are being conducted on the BeppoSAX mission \citep{fro98}.
The most complete catalogue of the orbital periodicities observed in the X-ray range as 
well as the transient sources was obtained as the result of observations during the Rossi XTE 
(RXTE) mission \citep{lev99,bra00a,bra00b}. 

	Although a large amount of observational data have been obtained by the present time, it 
seems that the experimental studies of periodicity as well as the other temporal phenomena 
(primarily outbursts) in X-ray sources have not yet exhausted their potential. One of the main 
problems is that not all of the observed periodicities can be unambiguously interpreted. In particular, 
what is the nature of the so-called non-eclipsing periodicities in the orbital range of periods? Several 
sources are known, which manifested periodic variations in the X-ray flux during which the X-ray 
luminosity does not fall to zero. Such a periodicity (8.2 h) was discovered in the emission from Cen 
X-4 during the well-known outburst of 1978, May \citep{kal80} observed in the Ariel-5 
mission. The dip-type periodicity was detected during the EXOSAT observations. It was observed in 
the X-rays from several low-mass binaries including MXB1759-29 (7.1 h), 4U1822-37 (5.6 h), 
4U1755-33 (4.4 h), X1323-62 (3.8 h), XB1254-69 (2.4 h), etc \citep{com84,mas81,whi84,par89,cou86}. 
Although periodic dips are not full eclipsing the common view is that all these periodicities are orbital. 
The 62-h periodicity in hard emission of GRO J1655-40 source also is not full eclipsing. Moreover in soft 
X-rays 62-h periodicity reveals itself as a number of deep drops in the intensity occurring between 
orbital phases 0.7 and 0.9 \citep{kuu98}. The light curve corresponding to this periodicity looks like a 
dip-type one with minima dislocated from the optical occultation. Thus, simple mechanisms such as 
a total or partial eclipse of the emitting hot region by the optical component do not give a complete 
explanation of the temporal behavior of the tight binary system's hard emission.
	Another problem, which still remain is the lack of data concerning the observation of 
periodic processes in the high-energy part of the electromagnetic spectrum. For most of the known 
tight binaries their orbital period values and corresponding light curves were obtained as the result 
of optical observations. On the contrary, only complete data from both optical and high-energy 
observations can give us the necessary possibility, which allows to determine the orbital parameters 
as well as the mass of binary system components. Thus, it seems important to find the manifestation of 
orbital periodicity in hard emission for the larger number of tight binaries.

	Search for periodicity in the emission from objects with great irregular (noise-type) flux 
variations or sporadic flux rises (transients) can also bring us closer to the understanding of the 
nature of such sources. The interest to the latter has increased in connection with the observations 
on the CGRO of the so-called transient pulsars \citep{fin96,zha96,wil97}
and sources with periodically recurring transient rises in the X-ray flux \citep{bay93,dek93}.
Study of temporal and spectral properties of sources known as X-
ray transients can clarify the nature of such objects and give us important information about high 
energy processes responsible for hard emission generation. From this point of view the search for 
periodic processes in hard emission during outbursts of sources known as X-ray novas is of 
particular interest. Most of X-ray binaries - black-hole candidates are just X-ray novas or transient 
sources \citep{sun91,cher97,van98}. Thus the study of temporal 
behavior of such objects can give us independent criteria which permit to identify black holes. For 
example, identification or, on the contrary, non-identification of pulsation-type periodic process 
during the outburst can clarify the nature of a compact companion, because the presence of 
pulsation excludes the black-hole variant. On the other hand, the spectral information, such as the 
presence of absorption lines at cyclotron frequencies, can also indicate the presence of a magnetic 
neutron star. On the contrary, the presence of an annihilation line is a strong indication of a black 
hole as the compact companion \citep{sun91}.
To the present time the information on transient X-ray sources was obtained primarily 
during their outbursts \citep{chen97}. However, the observations of such sources between outbursts 
can also be rather fruitful. There are some predictions that X-ray fluxes from transient sources even 
in quiescent state are non-zero. In particular, according to \citep{las00} the X-ray luminosity of 
quiescent black hole low-mass X-ray binary transients should be correlated with their orbital period. 
It is very unlikely that X-rays from such systems are emitted by coronae of companion stars. More 
convincing models associate their X-ray luminosity with accreting mechanisms, providing the flux 
generation at the observation level of such missions as Chandra or XMM. The measurements by 
Chandra showed that typical quiescent intensity of BH X-ray nova is about 10-6 Crab in 0.5-7 keV 
energy range. There are some indications that quiescent X-ray flux from BH X-ray nova is 
significantly lower than from NS ones and this fact can be caused by the existence of event horizon 
\citep{gar00}. Thus, the observations of X-ray transients in quiescent state can give unique 
information clarifying the nature of such objects.
Below we will discuss the results of the search and study of periodic processes in hard X-
rays from some Galactic sources including the black hole candidates during the observations which 
were carried out on the Prognoz-9 and "Mir" orbital station (OS) missions.

\section{MONITOR OBSERVATIONS OF TEMPORAL PHENOMENA DURING THE 
"PROGNOZ-9" MISSION AND "GRIF" EXPERIMENT ON BOARD "MIR" STATION}

\subsection{The "Prognoz 9" and "Mir" OS GRIF experiments.}

Both, Prognoz-9 and GRIF experiments use wide-field, hard X-ray spectrometers, which 
provide long-term observations of periodic process sources. Although the observational conditions 
were specific in the each experiment, the main X-ray instruments as well as the technique of the 
periodic sources revealing were quite similar.
The observations of galactic sources in hard X-rays (10-200 keV) were made in 1983-84 
during the complex experiment on a high-apogee ($\sim 720000$ km) satellite Prognoz-9 with a 
wide field of view (FOV) ($\sim 45^o$ FWHM) scintillator spectrometer ($\sim 40 cm^2$ effective 
area) \citep{kud85}. The X-ray instrument was installed in such a way that the 
centre of its field of view, averaged over the satellite's rotation period ($\sim 120 s$), coincided 
with the spin axis which pointed in the solar direction every 5-7 days. According to the experiment 
conditions sky areas adjacent to the ecliptic plane ($\pm 25^o$ - for instrument beam FWHM) were 
observed and slow ($1^o/day$ on average) scanning along the ecliptic was made. The count rates 
for X-ray photons were measured over the energy ranges of 10-50, 25-50, 50-100, 100-200 keV.
The region of the sky that was observed during the experiment is shown in equatorial 
coordinates in Fig 1a. The points of the sky toward which the satellite's axis was pointed at different 
times (the dates in the figure refer to the origin of the corresponding intervals of constant 
orientation) are also marked in this figure. The main black hole candidates and the Galactic equator 
are shown. In Fig 1b, the count rates (averaged over the intervals of constant orientation) in the X-
ray channels and in the charge-particle channel are plotted against time. The minimum count rates in 
the band 10-50 keV correspond to the time of observation when the Galactic-equator regions were 
outside the instrument's field of view. During the observations from November 1983 to February 
1984 of the sky region near the Galactic Centre (outlined by the closed solid line in Fig. 1a), the 
count rates were considerably higher than the background count rates; their maximum occurred in 
the time interval, during which the instrument was pointed nearly at the Galactic Centre region, 
where the Galactic X-ray sources concentration is the highest. Since each source in that region can 
be observed as long as 100 days, while virtually continuous measurements of count rates averaged 
over 10 s were made, the experiment provided favorable conditions for the study of periodic events 
over a wide range of periods.
The long-term observations of the Galactic Centre region as well as of some other sky 
regions were also conducted during the GRIF experiment on-board "Mir" orbital station (mean 
altitude $\sim 400$ km, orbit inclination $51^®$, orbital period $\sim 90$ min) from October, 1995 
to June, 1997 \citep{kud98a}. The scintillation spectrometer PX-2 for the energy range 
$\Delta E_{\gamma} = 10-300$ keV of detected photons with the effective area $S \sim 300 sm^2$, 
and field of view $\Omega \sim 1$ sr was the main instrument for astrophysical observations. It 
consists of 7 identical detector units of the "Prognoz-9" instrument type with crossed FOVs. The 
axes of these detector units were shifted on $5^o$ respectively to each other. This allowed us to 
observe almost the same area of the sky with all detectors simultaneously, and on the other hand, in 
the case of temporal phenomena registration to determine the source direction by the output data 
from each detector. The instrument provides flux measurements in energy ranges: 10-50, 25-50, 50-
100, 100-200 and 200-300 keV. Information was transmitted to the Earth in 16-h-long sessions of 
continuous observations; the interval between them typically ranged from several hours to several 
days. A total $\sim 200$ sessions were conducted during the experiment, from which $\sim 150$ 
without many telemetry failures and incorrect times of output data recordings were chosen for 
subsequent analysis.
Due to the rigid fastening of the detector units to the station instrument panel its orientation 
was determined by the station orientation mainly in two modes: 3-axes stabilization and orbital. In 
the first case the instrument axis had fixed orientation in space while in the second case it slowly 
($\sim 4^o/min$) scanned the sky by the station orbital motion. Thus the different parts of the sky 
including Galactic Centre and Anti-Centre regions were accessible for observations during this 
experiment. 

	The sky region observed during the experiment is shown in Fig. 2 in equatorial coordinates. 
Different shades of gray represent the exposure time throughout the entire experiment. This time 
corresponds to the points with respective coordinates. When estimating the exposure time, we 
assumed the angle between the source direction and the PX-2 axis to be no larger than $30^o$. In 
addition, we exclude the "Mir" residence time in the regions of trapped radiation. As a condition for 
the source being not shadowed by the Earth, we considered the requirement of PX-2 orientation to 
the sky, i.e. the angle between the PX-2 axis and the nadir-zenith direction (should be within the 
range $0^o-90^o$. The figure also shows the brightest X-ray sources and the Galactic equator. The 
typical exposure time can be determined using observing conditions for the Galactic Centre. 
Figure 2 shows a circumference with a radius of $30^o$ whose centre coincides with the Galactic 
Centre. There are sources within the region in the sky bounded by this circumference during the 
observations of which the PX-2 effective area accounted for no less than $50\%$ of its geometric 
area. As we see from the figure, the total observing time of the Galactic Centre with $\geq 50\%$ 
efficiency was $\sim 200$ h. The exposure times of other Galactic sources (for example, 4U1700-
37) are similar.

\subsection{The data processing technique}
	To reveal the periodic processes the time sets of X-ray instrument outputs were analyzed. In 
the case of the Prognoz 9 experiment such primary outputs were the mean count rates for 10 s. In 
the case of "Mir" GRIF experiment primary outputs were the mean count rates for 5 s. The data time 
sets formed by count rates averaged over longer time intervals (2 min, 10 min, 1 h) were also 
considered. These time sets were subject to random and regular variations, which produce the 
background for periodicities. These include rises in the X-ray flux due to solar flares and cosmic 
gamma-ray bursts, changes of the background count rate in the X-ray channels caused by variations 
of charged particle fluxes that bombarded the spacecraft, and some others. The counts in X-ray 
channels can be caused partially by the stochastic variations in the total intensity of the emission 
from the sources, which were simultaneously within the instrument's field of view. Both, Prognoz 9 
and "Mir" GRIF experiments were capable of eliminating the effect of some background factors 
with the use of direct measurements of the individual background components.

	Significant sporadic rises in the count rates in the X-ray channels were mainly attributed to 
hard X-ray bursts from solar flares and cosmic gamma-ray bursts. Since the latter were recorded 
during the experiments rather rarely \citep{kud88a,kud02}, 
they could not significantly affect the background in the search for periodicities. In the Prognoz 9 
experiment the Sun was constantly within instrument's field of view, thus its X-ray activity was 
monitored continuously, and about 800 solar X-ray bursts were detected \citep{abr88}. 
To identify such bursts the data from the Prognoz 9 RF experiment \citep{val79}, during 
which the Sun was continuously monitored in the band 2-10 keV by an instrument with a small 
($<10^o$) field of view, also we used the data of ground-based optical and radio observations of the 
Sun \citep{cof84} . In the search for periodicities the time intervals during which the 
solar bursts were detected were excluded from analysis. Since such bursts are rather short-lived, the 
total duration of the time intervals rejected in this way in Prognoz 9 experiment was short, no longer 
than 0.3
region. During the OS "Mir" GRIF experiment solar X-ray bursts were not detected at all because of 
the low solar activity during that time (1995-1997) \citep{cof03}.

	To remove those variations in the Prognoz 9 experiment outputs which correlated with the 
variations of the charged particle fluxe regression analysis of the count rates in the X-ray channels 
($N_x$) and in the charged-particle channel ($N_z$) was used. The relation between the count rate 
in a given X-ray channel and the count rate of charged particles detected by the anticoincidence cap 
can be assumed to be linear. Thus, the initial count rates in the analyzed time series can be 
represented as a superposition of the count rates that characterize the photon flux $\sim N_{x}^*$ 
under study and the additional count rates due to the charged particles ($\alpha N_z$):
\begin{equation}
N_x = N_{x}^{*} + \alpha N_{z}.
\end{equation}
The linear regression coefficients $\alpha$ were determined from the sufficiently long ($\sim 100$ 
days) time series that corresponded to the regions of the sky under consideration. The most 
significant linear regression coefficients were obtained for the channels 50-100 and 100- 200 keV 
($\alpha \sim 0.04$ photon per particle). The effect of variations in the flux of charged particles in 
the channels 10-50 and 25-50 keV turned out to be weak. The $\alpha$ coefficient values were used 
to obtain the time series of count rates $N_{x}^{*} = N_{x} - \alpha N_{z}$ for subsequent 
analysis.
	One of the main features of the GRIF experiment was the possibility of simultaneously 
monitoring all the principal components of background-producing emissions in the near-Earth space 
on "Mir" station orbits. Thus, the large-volume CsI(Tl) scintillator detectors of NEGA-1 instrument 
independently detected the local gamma-quanta ($\Delta E_{\gamma} = 0.15-50$ MeV, 
$S_{\gamma} \sim 250 sm^2$) and neutrons ($\Delta E_{n} > 20$ MeV, $S_{n} \sim 20 sm^2$) 
produced by the interaction of cosmic rays with the spacecraft material and the Earth's atmosphere. 
The FON-1 electron detector ($\Delta E_{e} = 40-500$ keV) of high sensitivity was used to check 
the sporadic increases in X-ray flux attributable to bremsstrahlung from precipitating energetic 
magnetosphere electrons, which could simulate astrophysical phenomena (gamma-ray bursts, 
transients). Due to a large geometric factor ($\Gamma \sim 80 sm^{2}sr$) it could detect even 
relatively small electron fluxes outside the zones of captured radiation. The FON-2 charged-particle 
detector ($\Delta E_{e} = 0.04-1.5$ MeV, $\Delta E_{p} = 2-200$ MeV), which was free from 
saturation in the radiation belts because of its small geometric factor ($\Gamma \sim 0.5 sm^2sr$), 
was used for background measurements when the "Mir" station crossed the South-Atlantic anomaly 
and spurs of the outer radiation belt.
	Since the relatively large orbital inclination and periodic crossings of the zones of captured 
radiation the PX-2 instrument background count rates underwent variations. However, the 
combination of active and passive shields greatly reduced the background variations, in particular, 
the latitudinal variations in the main X-ray channels 25-50 and 50-100 keV. To extend the 
sensitivity range of the instrument the latitudinal count rate variations  
were removed using regression analysis of the PX-2 X-ray 
channel outputs ($N_x$) and the NEGA-1 gamma-quanta channel outputs ($N_{\gamma}$). Since 
the NEGA-1 detectors were inside the OS "Mir" orbital module, they detected mainly the local 
gamma-emission. The additional count rate in a given X-ray channel attributable to the local 
emission may be assumed to depend linearly on the count rate of gamma-quanta detected by NEGA-
1. In this case, the initial count rates in the analyzed time series $N_x$ can be represented as a 
superposition of the X-ray count rate proper $N_{x}^*$, which characterizes the photon flux under 
study, and the additional count rate $\alpha N_{\gamma}$ attributable to the local gamma-
emission:
\begin{equation}
N_x = N_{x}^{*} + \alpha N_{\gamma}.
\end{equation}
The linear regression coefficients were determined over the entire observation interval when there 
were no bright sources of hard emission within the PX-2 field of view. The $\alpha$ coefficient 
values were used to obtain the time series of count rates $N_{x}^{*} = N_{x} - \alpha 
N_{\gamma}$ for the subsequent analysis. The suppression of background variations with the use 
of readings in the 150-500 keV gamma-quanta channel yielded the most significant result. After 
applying the regression procedure, the residual variations in the X-ray channels attributable to the 
latitudinal variations accounted for no more than $\sim 3\%$ of the corresponding means, which is 
several times less than the expected amplitude of the variations attributable to the emission from the 
most intense Galactic source.
	In the search for periodicities, the time series of count rates (which were cleaned from solar 
bursts and background variations caused by charged particles) were processed by the standard 
epoch-folding technique (see, e.g. \citep{ter92}, which was modified to accommodate the 
specific features of the Prognoz 9 and GRIF data. The intervals of observations were broken up into 
segments with duration equal to the trial period under consideration. The sequences of count rates 
that corresponded to these segments were added together, and an average phase dependence of the 
count rate was constructed for this trial period. The amplitude of the periodicity corresponding to 
the trial period (the actual or randomly simulated one) can be described by the rms deviation 
(${\sigma}^2$) of the numbers $M_i$ that constitute the average phase dependence:
\begin{equation}
\sigma^2 ={\sum\limits_{i=1}^K (M_i - \bar{M})^2 \over {K-1}}.
\end{equation}
Here $\bar{M}$ is the mean count rate determined from the entire analyzed time series, $K = {T \over 
\Delta T}$, where $\Delta T$ is the bin duration of the count rates that constitute the mean phase 
profile; and $T$ is the trial period. If the periodicity related to the period under study contains both 
a real periodic component (pc) and a noise component, then because of their independence, ${\sigma}^2$ 
can be represented as the superposition:
\begin{equation}
\sigma^2 = \sigma^{2}_{pc} + \sigma^{2}_{noise},
\end{equation}
where $\sigma^{2}_{pc}$ and $\sigma^{2}_{noise}$ characterize the amplitudes of the 
corresponding components. 
In general, the dependence $\sigma^{2}(T)$ (periodogram) can be represented as the 
superposition of a noise continuum (ideally, a smooth function of $T$) and discrete peaks of the 
existing periodicities that correspond to the main period and its multiplies. Figure 3a shows an 
example of the periodigram, obtained from the Prognoz 9 data at the fixed bin length of 1 h 
for the time series of count rates in the channel 25-50 keV corresponding to the interval of 
observations of the Galactic Centre region (October 31, 1983 - January 12, 1984). The periodogram 
clearly shows the discrete peaks that belong to the 82 h periodicity: the main period, half-period, 
and its multiplies (82 h is the orbital period of the X-ray binary 4U1700-37, which was constantly 
within the instrument's field of view during the interval of observations). In this representation, the 
continuum in the periodogram is a monotonic, virtually linear increase (a trend) in 
$\sigma^{2}_{noise}$. For a purely harmonic periodicity, the peak in the periodogram has a finite 
FWHM $\Delta T_{source} \approx {0.9T_{source}^2 \over T_{max}}$, where $T_{source}$ is the period of 
source, $T_{max}$ is the total time of observations. It should be noted that, in real cases, the relative 
FWHM of the peak can be larger (if the periodic signal was recorded during a time interval shorter 
than the duration of the entire period of observations under analysis) and smaller (if the mean phase 
profiles differs markedly in shape from a harmonic profile, for example, has the shape of narrow 
dips). In the absence of additional information, the FWHM of the peak, which is determined directly 
from the periodogram analysis, can serve as the measure of an error in the period. 
In general, the continuum in periodograms is determined by a superposition of different 
(noise) components. One of such noise factors, shot noise ($\sigma^{2}_{shot}$), is related to the finite 
number of photons with a Poisson distribution recorded per bin. This component in the periodogram 
can be calculated from the mean count rates in the channels; it is linear for periodograms with a 
constant bin length, 
\begin{equation}
\sigma^{2}_{shot}(T) = \bar{N} \cdot {1\over\Delta} \cdot {T \over T_{max}},
\end{equation}
where $ \bar{N} $ is the mean count rate, $ \Delta $ is the bin length. Clearly, this component is always present in 
the noise, and the observed continuum can not be lower. The excess over the "shot" component in 
the total noise continuum can be explained by non-periodic variations in the fluxes from the 
observed X-ray sources (Galactic noise) and by variations in the instrumental background. 
For a wide range of periods, the results of data processing by the epoch-folding technique 
are more conveniently presented by plotting the inverse period (frequency) along the x-axis and by 
specifying a linear frequency grid in the calculations. In this case, the FWHM of the peaks 
corresponding to periodicities with similar shapes is the same in the frequency spectrum at any 
periods. If we plot the parameter ${\sigma^2} \over K$ ($K$ is the number of bins in the trial period) along the $y$-
axis, the frequency continua of noise components correspond, to an accuracy of a factor, to those of 
the power spectrum in a Fourier analysis. In particular, the quantities that characterize shot noise are 
constant at all frequencies (white noise). The scatter of points in the real spectrum about the 
frequency-averaged values of ${\sigma^2} \over K$ in the noise segments in the vicinity of the peak 
under consideration gives the variance $\sigma_{\sigma}$, which can be used to estimate the 
significance of the periodicity corresponding to this peak (in the case of Prognoz 9 data 
$\sigma_{\sigma}$, values were calculated with the use of points
in the frequency range $\pm100 \cdot FWHM$ relative to the peak; for 
these intervals, the background level between peaks was essentially constant). Since the determining 
of the mean values of ${\sigma^2} \over K$ was performed by averaging over a large number of 
points in the spectrum, the error of the mean itself can be ignored, while the significance of the 
identification of a periodicity is determined by the ratio ${\sigma_p}/\sigma_{\sigma}$.
A comparative analysis of the data obtained in independent observation intervals can 
provide additional information on the significance of the detected periodicities. In particular, the 
phase profiles that were constructed by individually averaging the data corresponding to only "odd" 
and "even" time segments are independent. Clearly, these phase profiles placed one after the other 
constitute the mean phase dependence that corresponds to a doubled period. The correlation 
between features in the odd and even mean phase profiles confirms the significance of the 
periodicity itself and suggests that the time structure of the mean profile is real. Thus, the similarity 
of the even and odd mean phase profiles can be chosen as the additional significance criterion for 
weak periodocities. The correlation coefficient $k$ between the sequences of numbers constituting 
the even and odd profiles is a quantitative parameter that characterizes this similarity. 

\section{RESULTS OF OBSERVATIONS IN THE PROGNOZ 9 AND GRIF EXPERIMENTS}

\subsection{The Prognoz 9 summary catalogue of periodicities}

Using the method described above, all the data that were obtained during the Prognoz 9 X-
ray experiment were analyzed. To select the significant periodicities the condition that the amplitude 
of the peak corresponding to the main period should exceed the mean noise value 
${\sigma^2}_{noise} \over K$ by more than $5\sigma_{\sigma}$ was established. Under the 
condition of a normal (Gaussian) distribution of the noise values of ${\sigma^2} \over K$ about the 
means, this criterion provides a selection probability of a single "random" peak at a $1\%$ level in 
the search for peiodicities at all frequencies and over the entire interval of observations of the 
Galactic Centre. Apart from the requirement $>5 \sigma_{\sigma}$, the full significance criterion 
for a periodicity was supplemented with the condition $k \geq 0.5$ (for 40 bins in the phase profile). 
The periodicities were sought in the range of periods from 0.5 to 200 h. The result is the number  
of periodicities in hard ($kT \geq 40$ keV) X-ray sources: 152, 98, 
82, 69, 62, 13, 9.4, 8.15, 8.04, 4.38, 4.35, 3.45 h \citep{kud88b,kud92,kud98b}. 

A list of revealed periodicities and some of their parameters are given in the table~\ref{tbl-1}. It 
should be noted that the survey depth (minimum average fluxes of periodicities identified above the 
background) was slightly different for different regions of the sky and in different ranges of periods. 
The maximum noise level was determined by random variations in the total emission from the 
sources near the Galactic Centre. For this region, the 25-50 keV fluxes of periodicities could be 
recorded starting from levels $\geq 2.5 \cdot 10^{-3} phot/cm^2 \cdot s$, on the average, for $T = 1 - 
30$ h and $\geq 5.0 \cdot 10^{-3} phot/cm^2 \cdot s$ for $T = \geq 30$ h. Although the sensitivity in 
the survey of sky regions that lay far from the Galactic Centre was higher, the revealed periodicities 
all belong to the Galactic-Centre region and adjacent regions. This is in good agreement with 
standard spatial distribution of X-ray sources in general (not necessarily with periodicities).
The values of periods were determined by the maximums of the peaks in $\sigma^2 \over K$ 
versus period periodogramms. The resolution on revealed periods defined by the peak width 
depends on the ratio of period value and the time corresponding to periodicity observation. This 
resolution changes from $\sim 10\%$ (82 h, 98 h, 152 h) to $\sim 1\%$ (8.0 h). To identify some of 
these periodicities we used information about period, flux, spectrum and location of the possible 
source. The 82 h period was identified naturally with orbital period of the eclipsing binary 4U1700-
37 \citep{kud91}. The 8.2 h period was discovered by \citep{kal80} 
in Cen X-4 X-rays during the outburst in May 1979. The 8.2 h periodic process was observed in the 
Prognoz 9 experiment during the limited time interval when except for Cen X-4 there were no 
sources with similar periodicities in X-rays in the instrument FOV. Hence, we suppose that this 
period can be connected just with Cen X-4 although we have no information about the brightness of 
Cen X-4 at that time \citep{kud95}. The 4.35 h periodicity was identified in the same 
way with the equivalent period of dips in 4U1755-33 X-ray flux \citep{whi84}. The other 
periodicities mentioned above were hitherto not observed. The values of the effective temperature 
kT (for thermal bremsstrahlung spectra) were evaluated for each periodic process. This parameter 
characterizes the spectral hardness of the periodic flux component (only in the case of eclipsing 
binary 4U1700-37 it corresponds to the full flux spectrum).

\subsection{Location of the periodic sources observed by Prognoz 9}

Due to the wide field of view of the X-Ray instrument it was impossible to locate with 
sufficient accuracy the sources of periodicities revealed in the Prognoz 9 experiment. However, in 
many cases it was possible to estimate the error box of source location. The detector beam was 
isotropic in azimuth and linearly fell from the peak ($\theta = 0^o$) to zero ($\theta = 45^o$) with 
increasing offset angle. The instrument's FOV scanned the sky, when the satellite moved with the 
Earth around the Sun. Then for every source in the instrument's FOV there were the moments of 
time, when the source only begun to be seen, when it became visible under the favorable conditions 
(i.e. the angle between direction on the source and the instrument's axis was minimal), and at least, 
when the source left the instrument's FOV. Because at the moment of best visibility of the 
source, its offset angle was minimal, the source should be located on the line of a big circle, which 
is normal to the ecliptic plane and contains the point of the ecliptic, in which the instrument's axis 
was directed in that time. The full time of the visibility of the source depends on the minimum 
offset angle, which can be achieved during the scanning. If the source was caught by only the very 
edge of the instrument's FOV, then the time of it visibility was very short. In the opposite case, 
when the source during the scanning arrives near the centre of the instrument's FOV (it was the 
case of the Galactic Centre) and it luminosity was constant, we have the maximum full time of it 
visibility, i.e. about 90 days. It means that, if the source was located near the instrument axis 
trajectory, amplitude of the corresponding peak on the periodogramm should not be less than half of 
its maximum value during a month and a half at least. The objects located in such a way, that their 
offset angle $\theta$ is $\sim 30-40^o$, should be visible during more than a three weeks at the level of 
about $1/4$ of maximum intensity at the moment of their best visibility. If a periodic process 
was observed during only one or two intervals during which the satellite orientation was constant, it 
should be supposed that source intensity variations were caused not by the changes in the 
instrument's beam orientation, but due to the variability of the source itself.
If the source was unknown, it was impossible to determine which part of the total flux, which 
is detected in the time of the best visibility. However, if we assume the constant luminosity of a 
source, it is possible to evaluate the time of the source visibility using information on the 
intensity of the flux periodic component at different times, which should be normalized to the 
maximum observed intensity. Then, if the observation time of the maximum periodic 
component intensity is known, it is possible to obtain the offset angle of the source and, thus to 
locate it in the sky with an error box much smaller than the instrument's FOV. Nevertheless, the 
size of such error boxes (about $10^o-20^o$) makes it very difficult to identify clearly the sources 
of periodic processes, particularly in the case of the regions adjacent to the Galactic Centre, where 
the number of X-ray objects - possible candidates of the sources of observed periodicities is rather 
high. The identification can be much better in the case, when the source of intensive periodic process 
is located sufficiently far from the Galactic Centre. In such cases even despite a wide error box  
the number of objects - possible candidates for the periodic sources issufficiently small.
Thus, by location of the sources of observed periodicities, an individual approach was used 
in every case. To make identification better we use not only the data from Prognoz 9 and GRIF 
experiments, but also the data of some other well-known X- and gamma-ray observatories. We 
will discuss further, the indications that the sources of 152 h and 62 h periodicities can be identified 
with X-ray Novas Ophiuchi, 1977 (H1705-25) and Scorpius, 1994 (GRO J 1655-40) 
correspondingly. The 98 and 69 h periodicities were observed during all the time when the Galactic 
Centre was in the instrument's FOV. Thus, we can assume that their sources are positioned near the 
Galactic Centre, but they could not be located with accuracy better than FOV.
Among the periodic processes with hour range of periods, at least three (13.3, 9.36, 8.04 h) 
were observed during the time interval which was sufficiently long to make possible the location of 
their sources. The part of the sky with error boxes for possible sources of 13.3, 9.36, 8.04 h 
periodicities is shown in Fig. 4. The regions of the source location were obtained for the case of 
their constant luminosity. The error box edges correspond to a $70\%$ ($1\sigma$) probability of 
the source location within the box. The sources with luminosity sufficient to be observable in the 
Prognoz 9 experiment are also marked in the Figure. These sources include objects from the 
HEAO-1 A4 catalogue \citep{lev84}, whose fluxes in the range 13-80 keV exceeded 
$2.5\cdot 10^{-3} phot/cm^2s$. This value corresponds to the Prognoz 9 X-ray instrument sensitivity 
level for the detection of periodic processes. This catalogue includes not only steady state sources, 
but transient pulsars also, because the typical time between their outbursts was much less than the 
time of the experiment. The X-rays novas are also shown in Fig. 4. Their location is given according 
to \citep{chen97}. Because of the high intensity of X-ray novae outbursts, during which the 
flux can increase up to 100 mCrab, they can be observed even in the case of large angular distances 
from the centre of the instrument FOV. Typical duration of the outbursts is about 10-300 days, 
permitting us to assume a quasi-constant character of their fluxes during observations.
As it can be seen from Fig. 4 the error boxes for the sources of 13.3 and 9.36 h periodicities 
are located rather far from the Galactic Centre. Below we will discuss the indications on that the 
source of 13.3 h periodicity is the X-ray transient - black hole candidate 4U1543-47.
The 9.36-h periodic process was observed during the time interval including about 70 full 
periods. The error box bounding the location region of the source contains only a few well-known 
objects. However, now we can not give a definite identification of the source of this periodicity.
In the case of the 8.04-h periodicity the source may be associated with many objects located 
along the Galactic Plane. The source of this periodicity was seen in the instrument's FOV during a 
long time, indicating that the object is positioned no further than $ 20^{o} $ from the instrument axis 
trajectory ant its luminosity is quasi-constant. If we also take into account the position of the 
periodic process intensity maximum relatively to the instrument axis trajectory (see Fig. 4), we can 
conclude that the source of 8.04-h periodicity lies in the Galactic bulge.
The source of 4.38-h periodicity was observed during approximately two weeks (from 
28.10.83 to 10.11.83). This time interval contains more than 70 full periods. The relatively short 
time of this periodic process visibility can be explained by two reasons: either rather high variability 
of its source or the source was visible near the edge of instrument's FOV. If we assume quasi-
constant intensity of the 4.38-h process, then its source can be located in the sky region where well-
known sources 4U1439-39, H1409-45 are positioned. The alternative is the very bright source of X-
ray nova type located in the direction of the source GS1354-64 (see Fig.4).
The periodic process with 3.45 h period was observed most vividly during one interval of 
the instrument axis constant orientation in the direction of Galactic Centre. During the time of the 
next orientation it was not visible. This indicates that the source of this periodicity could be one of 
the high-variable sources near the Galactic Centre.

\subsection{The mean light curves of the periodic sources observed by Prognoz 9}

The lines in Figure 5 present the mean light curves of long-term periodic sources. The 
mean line curves, which correspond to the periodicities of hour range of periods are shown in 
Figures 6. The mean light curves in different energy ranges are presented on the left panels of 
Figures 5,6. To confirm the reliability of the revealed periodicities themselves and to distinguish 
between the temporal (stochastic) and phase peculiarities (peaks) in the mean light curves, we 
considered besides the mean phase profiles corresponding to the main periods the mean phase 
profiles corresponding to the "double" periods. The latter ones were divided into two "halves". 
From the first "halves" the "odd" profiles (upper dotted lines on the right panels of Figures 5,6) 
and from the second "halves" - the "even" ones (lower dotted lines on the right panels of Figures 
5,6) were composed. Thus, we folded separately data for non-overlapping time subsets with the same 
reference point. The similarity of "even" and "odd" phase profile forms confirms the reliability of 
periodic processes. 
It is necessary to note that except for the 82 h light curve all other curves are not typically 
eclipsing. Because most of the small details for mean light curves can be caused by temporal 
stochastic variability, we can definitely conclude that they correspond to complicated phase 
structure only if such details are present on both "odd" and "even" profiles. Such "repeated" details 
can be seen on light curves corresponding to 8.04 h, 62 h, 82 h and 152 h periodicities.
The quasi-symmetrical structure of 82 h light curve consisting of the 4 local maximums (at 
$\sim 0.2, \sim 0.3$ and $\sim 0.7, \sim 0.8$ of the full phase), which can also be seen on non-
overlapping observation intervals, should be especially noted. Since this light curve was obtained as 
the result of continuous observations of X-ray binary 4U1700-37 during the time interval containing 
8 full orbital cycles, its structure can reflect the peculiarities of compact companion orbital motion.
The 8.04-h periodic process is characterized by a mean phase profile with two well-
pronounced peaks. This two-hump structure is confirmed by the profiles in the different energy 
ranges: 10-50, 25-50, 50-100 keV. From this we can conclude that emission in the narrow peak is 
much harder, indicating the existence of regions with different temperature in the source. This is 
similar to the source of 152-h periodicity \citep{kud96}. 
The 13.3-h periodicity is characterized by the mean phase profile with a deep minimum, 
which width is about $25-30\%$ of the period. Near the maximum intensity the light curve can be 
presented as the superposition of a symmetric smooth component and the quasi-periodic fast variable 
component. Fast variability can be seen on both the "odd", and the "even" profiles. However, the 
sub-harmonic of about 1.8-h at 13-h light curve is not confirmed because it is clearly revealed only 
on the "odd" profile. As it can be seen from Figures 5c, 5d the mean phase profiles of 9.36 h and 
4.38 h periodicities are quasi-harmonic, while the phase profiles of 8.15-h (Cen X-4), 4.35-h 
(4U1755-33) and 3.45-h periodic processes have a dip-like form.

\subsection{Results of periodic source observations conducted during the GRIF experiment on-board "Mir" 
orbital station.}

	For the search of periodic processes in the GRIF experimental data time sets of 10-min 
averaged X-ray count rates in various energy ranges (which were cleared of background variations 
attributable to latitudinal variations) were used. To reveal periodicities by the epoch-folding 
technique, these time series were formed for intervals of observation of a particular source chosen 
according to the criteria, which were considered above in the section concerning the observational 
conditions in the GRIF experiment. The epoch-folding technique was modified taking into account 
the peculiarities of the GRIF experimental data, i.e. by obtaining a periodogram for each trial period, 
the average phase profiles broken down into 40 independent bins were calculated. For example, at 
an exposure time of $\sim 200$ h for 10-min averaging of the count rates that constituted the initial 
time series, there were, on average, 30 independent count rates in each bin.
The capabilities of the GRIF experiment to observe periodic processes in galactic sources 
can be estimated basing on the RX-2 instrument's background count and exposure time. The typical 
value of the latter can be characterized for example by the total exposure time of the Galactic 
Centre, which was about 200 hours. With the background counts $\sim2.5 \cdot 10^{-1} 
pulses/cm^2s, \sim 1.2 \cdot 10^{-1} pulses/cm^2s$ in the 25-50 keV and 50-100 keV channels, 
respectively, it corresponds to the $5\sigma$ level of minimum detected fluxes for a periodic component 
with period in the ranges of hours and days ($\geq 7$ h). In the case of a purely statistical (Poisson) 
distribution of count rates this level is $\sim 4 \cdot 10^{-5} phot/cm^2s$ in the 25-50 keV range 
($\sim 10$ mCrab) and $\sim 10^{-5} phot/cm^2s$ in the 50-100 keV range ($\sim 5$ mCrab). 
Figure 3b shows the periodogram obtained by the above technique from the GRIF 
experimental data for the time series of the count rates in the 25-50 keV channel selected for 
observing intervals in 1995 - 1997. During this time the PX-2 instrument axis was oriented to the 
sky region with the well-known sources of hard emission 4U1700-37 and GRO J 1655-40. The 
condition for choosing these intervals of observation was the requirement that offset of the PX-2 
axis from the direction of 4U1700-37 be no larger than $30^®$ (since the angular positions of 
4U1700-37 and GRO J 1655-40 are close in the sky, the time series composed for the latter under 
the same condition for the PX-2 axis offset virtually coincided with the time series for 4U1700-37). 
The periodogram exhibits distinct discrete peaks that correspond to periods of 62 and 82 h. A 
number of broad peaks corresponding to periods of 24, 48 and 72 h, which are attributable to the 
diurnal variations of the instrumental background in the PX-2 channels, are also seen on the 
periodogram. The significance of the separated peaks is specified by the dispersion 
$\sigma_{\sigma}$, which characterizes the scatter of points on the noise portions of the continuum 
near the analyzed peak. The noise continuum on the periodogram is determined by a set of various 
factors, which were discussed in previous sections. These include shot noise, attributable to a 
finite number of detected photons with a Poisson distribution, non-periodic variations on the flux 
from the source within PX-2 FOV; and variations in the instrumental background. The amplitudes 
of the peaks marked in Fig. 3b are 5.5$\sigma_{\sigma}$ (62 h), 8$\sigma_{\sigma}$ (82 h). If we 
assume a normal (Gaussian) distribution of the noise values of $\sigma^2(T)$, this leads us to 
conclude that the corresponding periodicities are statistically significant \citep{kud01}. 
As it was pointed out above, the 82- and 62-h periodicities were previously observed in the 
hard emission during the Prognoz 9 experiment. The 82-h period is the orbital period of the 
eclipsing X-ray binary 4U1700-37, which is known from several space experiments: UHURU 
\citep{jon73}, OSO-8 \citep{dol80}, EXOSAT \citep{doll87}, BATSE 
CGRO \citep{rub96}, SIGMA Granat \citep{sit93}, and others. The 62-h period is 
equal, within the error limits, to the optical orbital period of the X-ray binary GRO J 1655-40 
\citep{bai95}, which is known as the X-ray Nova Scorpii 1994 \citep{zha94}. Below 
we will discuss in details the indications on that 62-h periodicity can really be identified with this 
object.
Basing on the time series of count rates in the PX-2 10-50, 25-50, 50-100 keV channels, the 
periodograms for the observing intervals when the PX-2 axis was oriented to the Galactic Centre 
and to the region of H1705-25 (Nova Ophiuchi 1977) location were obtained by the epoch-folding 
technique. As was noted above, during the Prognoz 9 observations, the 98- and 69-h periodicities 
were revealed in hard emission from the Galactic Centre region. Some indications were obtained by 
\citep{kud96} that 152-h periodicity can be identified with H1705-25 (below we will 
discuss some details of this identification). However, the GRIF periodograms for the above PX-2 
orientation revealed no significant peaks corresponding to the 98, 69 and 152 h periods. We cannot 
rule out the possibility that the corresponding exposure time ($\leq 200$ h) for the 152-h periodicity 
is not sufficient for more or less reliable separation by the epoch-folding technique. The estimate 
upper limits on the spectral flux densities averaged over the 10-50, 25-50, and 50-100 keV energy 
ranges, which characterize the intensities of the 69-, 98- and 152-h periodicities, show that they 
cannot exceed the values given in table~\ref{tbl-2}, within $1 \sigma$.

Thus, by comparing the results of the searches for periodic sources during the Prognoz 9 and 
GRIF "Mir" experiments, we conclude that among those day-range periodic processes, which were 
detected in 1983 - 1984, the 62-h and 82-h periodicities continued to be observed in 1995 - 1997. 
The fact that 69, 98 and 152 h periodic processes were not observed in 1995 - 1997 can be 
explained by the variable nature of hard emission of those sources. As for the periodicities with hour 
range of periods, because of the high indefinites in location of their sources and "dangerous" 
nearness of those periods to the orbital period of the "Mir" station (1.5 h) and its harmonics, we do 
not hope to reveal them in the GRIF experimental data.

\section{IDENTIFICATION WITH R-RAY NOVAES - BLACK HOLE CANDIDATES}

\subsection{H1705-25 (XN Oph 1977) - the source of 152 h periodicity.}

Basing on all the data on 152 h periodicity obtained in the Prognoz 9 experiment, we 
concluded that it can be identified most probably with H1705-25 (XN Oph 1977) \citep{kud96}.
Since we know there were no other observations of this source in hard X-rays 
in 1983, November - 1984, January, we assume that it could be active at that time.
The source of 152-h periodicity is the most hard and long-periodic between the bright 
objects observed in Prognoz 9 experiment. The periodic component flux value at the energies 
$\sim100$ keV is no less than 100 mCrab. Only two objects in the error box of this source position 
can be characterized by such fluxes. They are the X-ray novas of 1977 (H1705-25 from HEAO 1 A4 
catalogue \citep{cook84} and 1993. The curves, which are presented in Fig. 6, confirm that the 
152 h periodicity can be associated with Nov Oph 1977. These phase dependencies were obtained 
by means of the epoch-folding technique. Then the SSI Ariel 5 initial data (the flux values J in 2-18 
keV energy range were averaged over $\sim 1.7$ h time intervals) were normalized by the averaged 
flux values J`, which present only the trend-like intensity monotonous decrease during the transient 
decline phase.
The value 167 h is equal to the 6.7 days period of quasi-periodic component in Nov Oph 
1977 2-18 keV X-rays, which was suggested by the authors of SSI experiment on the basis of their 
own experiment data analysis. The outstanding similarity of the curves in Figure 6 (especially in 
view of its complicated structure) should be noted. Because the difference of $152 (\pm 6)$ h and 
167 h period values exceeds the errors, it can be concluded, that during the time between both 
experiments (more than 5 years) a $\sim 10\%$ period drift took place.

\subsection{GRO J1655-40 (XN Sco 1994) - the source of 62 h periodicity.}

Due to the triangle-like form of the instrument beam the error box of the observed periodic 
source positions could be determined in some cases using the time of the full visibility of the source 
and the correlation of its amplitude with pointing direction. We have analyzed the data on X-ray 
sources (including X-ray novas) in Scorpius and the nearest constellations. As the result of this 
analysis we have obtained that 62 h periodicity can be identified with GRO J1655-40 (XN Sco 
1994) - an eclipsing binary with an orbital period of 2.62 days which is equal to 62 h in error 
limits \citep{kud98b}. 

	The CGRO observations revealed the transient source GRO J1655-40 in Scorpius, also 
known as the X-ray nova XN Sco 1994 \citep{zha94}. The orbital period of this system was 
estimated from optical observations to be 2.62 days = 62.9 h \citep{bai95}, which is equal, 
within the error limits, to the 62-h period observed in 1983 during the Prognoz-9 mission. The hard 
X-ray observations of this source also show significant variations in the X-ray flux in which, as was 
noted by the authors themselves, a quasi-periodicity with a similar period can be identified. The 
Prognoz 9 and GRIF mean light curves for the 62 h period are shown in Fig. 7. The light curve that 
was obtained by averaging the OSSE CGRO observational data for GRO J1655-40 \citep{kro96}
over the 62-h period is shown in Fig. 8. The Prognoz 9 and CGRO light curves were brought 
in coincidence in phase based on the maximum of the correlation coefficient between the 10-50 keV 
(Prognoz 9) and 50-100 keV (OSSE/CGRO) light curves, which was 0.76 \citep{kud98b}. 
The GRIF and Prognoz 9 curves were brought into coincidence by the middles of the phases 
of intensity minimums. As it can be seen from the Figure, the 
shapes of the all presented light curves do not contradict each other. The values of minimum phase 
duration are almost equal, $\sim 10\%$ of the period, and the intensities of the 62-h periodicity 
measured in these three experiments can be considered to agree with one another, at least to within 
an order of magnitude.
It is of great interest to compare the Prognoz 9 and GRIF observations of the GRO J1655-
40 activity with observations made by some other observatories. The activity of this object in its 
hard emission, which has subsequently become known as the X-ray Nova Sco 1994 \citep{zha94,har95}, 
was apparently first observed during the Prognoz 9 experiment in 1983 \citep{kud98b}.
According to the CGRO data, this source was most active in the hard 
energy range from July 1994 through August 1995 \citep{kro96,zha97}. 
Subsequently, late in April 1996, an outburst was again detected first in the soft X-ray (2-10 keV) 
range with the All Sky Monitor (ASM) of the RXTE Observatory and, about a month later, in the 
hard energy range at the Compton Gamma-ray Observatory \citep{hyn98}. The source 
continued to be more or less active for at least 16 months \citep{rem99}, and it was 
observed by various spacecraft, including RXTE, CGRO, BeppoSAX. The BATSE CGRO data 
suggest that the activity in the source's hard emission between outbursts was characterized by fluxes 
$\sim (3-5) \cdot 10^{-2} phot/cm^2 \cdot s$ in the 20-200 keV energy range. The intensity of the 62-
h periodicity in the hard emission from GRO J1655-40 was estimated from the GRIF data for the 
entire period of observation (1995-1997). The obtained value is $\sim (4.5 \pm 1.85) \cdot 10^{-2} 
phot/cm^2 \cdot s$ in the 25-200 keV energy range, which agrees with the BATSE CGRO upper limit 
on the total flux, within the error limits. The duration of the phase of minimum on the Prognoz 9, 
GRIF and OSSE CGRO average phase profiles for the hard emission is similar (see Fig. 7). 
The analysis of the RXTE ASM data, which are freely accessible on the RXTE Internet site 
\footnote{Center for Space Research at the Massachusetts Institute of Technology. Electronic Version is available at: http://xte.mit.edu}) also indicates the 
presence of the 62-h orbital periodicity in the 2-12 keV emission during the 1996-1997 outbursts. 
The RXTE ASM data were processed using the same epoch folding technique as in the processing 
of the Prognoz 9 and "Mir" GRIF data. The RXTE site interface permits to obtain the time set of the 
ASM outputs, which corresponds to one of the observed sources for the chosen interval of 
observations. The ASM data are accessible for observations beginning in 1996 (TJD 50088-52719). 
These data contain for each time of observation the source intensity as in the full instrument energy 
range 1.5-12 keV, as in the sub-ranges 1.5-3, 2-5, 5-12 keV. As a rule, the presented values of the 
emission flux were measured during about one 1.5-min time interval on each satellite orbit (the 
orbital period $\sim 90$ min).
The main peculiarity of the RXTE ASM output data is the rather short time of an individual 
observation of a source on each satellite orbit. This permits to consider such measurement as a 
point-like on the time interval. Thus, to use the standard epoch-folding technique, the processing 
time set was prepared in such a way, that each trial period was divided onto 20 independent 
intervals, i.e. parts of a full phase or bins. Each ASM output was attributed to the phase of a given 
trial period depending of the time of measuring. Then the intensity value was attributed to the 
corresponding bin of the phase curve. For each bin the intensity values were summarised and, 
hence, the mean phase curve were obtained. To estimate the possible periodic process amplitude the 
rms deviation of the values in separate bins from its mean arithmetical value (i.e. from the average 
source intensity) was calculated. Corresponding periodograms were obtained for the range of trial 
periods from the satellite orbital period (about 1.5 h) up to several hundreds of hours. The grid of 
the trial period values uniforme in frequency was used. In the case of an outburst the slow varying 
component of light curve should be subtracted from the primary data set.
The data set containing the part of the ASM RXTE output data for the time of the outburst 
activity of GRO J1655-40 in 1996-1997 was chosen for searches of 62-h orbital periodicity. The 
time intervals, for which the rapid variability of this source was relatively more intensive than the 
slow-changing component, were prepared for the analysis by epoch-folding technique. These time 
intervals are TJD 50138-50198 (before outbursts), TJD 50250-50350, 50500-50580, 50700-52200 
(flat intervals of outbursts) and 50700-51000 (after outbursts). The 62.9-h peak intensities on the 
corresponding periodograms are revealed quite significantly up to the level of $\sim 6.5 \sigma$. They 
are more vivid for the stages of quasi-stable emission of GRO J1655-40 during outburst: the best 
one is for 50500-50580 interval. 
The practically simultaneous observations of GRO J1655-40 in the GRIF "Mir" and ASM 
RXTE experiments allow us to make the phase attachment of light curves in different energy ranges 
and to compare the mean light curves. Since we know with high accuracy the time of observations it 
was possible to construct the mean light curves obtained in both experiments in one phase. These 
light curves as well as the mean phase profiles obtained from OSSE CGRO and
Prognoz 9 data are plotted on Fig. 7. As it can be seen from the Figure, 
the phases of minimum intensity or dip-phases practically coincide in the 2-12 keV (ASM RXTE) and 25-50 keV (GRIF) 
ranges. The practically equal time of the dip-phases for these energy ranges should also be noted. 
This fact is of special importance in view that the phase of minimum intensity as soft X-rays as hard 
X-rays and gamma rays does not coincide with the minimum of its optical brightness.

\subsection{4U1543-47 - the source of 13.3 h periodicity.}

Some indications on that 13.3 h period can be identified with X-ray transient source 4U1543-47 
known as X-ray nova - black hole candidate \citep{kit84,oro98} were 
obtained as the result of Prognoz 9 data analysis. The orbital period which was determined by the 
displacement of the lines in the optical spectrum of 4U1543-47 is equal $26.9\pm 0.2$ h, which in 
the error box limits coincides with the double period of 13.3 h. Despite that location accuracy in the 
Prognoz 9 experiment with wide-field instrument was not high ($\sim 10^o$), due to that field of the 
possible location of the source of 13.3-h periodicity is rather far from the Galactic Centre the object 
4U1543-47 was practically one between the other sources of hard emission in the error box. It 
should be also noted that 3 months before the observations on the Prognoz 9 in 1983 the beginning 
of rather bright outburst in this source was detected. Thus we may conclude that it could be active 
during the observations on Prognoz 9.
	The search of the orbital periodicity in the hard emission of 4U1543-47 was also made with 
the use of ASM RXTE data for time intervals:
A) TJD 50088-52440.
During this time source was in quiescent state. The corresponding time set consists of 
20600 bins for more than 6 years of observations. The range of ASM output counts 
was $\pm 10 s^{-1}$ that was caused, probably, by the background variations. 
B) TJD 52445-52466
This time interval corresponds to the outburst decrease, which was preceded by a fast rise of 
flux during about 4 days. During 21 days the flux was measured 187 times. The count rate of 
4U1543-47 emission detected by the ASM instrument in the range 1.5-12 keV was changed from 
$330 s^{-1}$ to $40 s^{-1}$. For further analysis by the epoch-folding technique, the quasi-
exponential slow component was subtracted from the primary time set. To approximate this slow 
component the method of sliding linear approximation was used. 
	') TJD 52510-52719
This time interval corresponds to the quiet state after the outburst. 
There are no significant peaks on the periodograms obtained for the quiet state of source (A and 
C). However, on the periodogram obtained for the outburst decrease time interval (B) peaks 
corresponding to $\sim 27$ h and $\sim 13.5$ h periods were revealed at the levels $\sim 5 \sigma$ 
and $\sim 6.5 \sigma$ respectively. The probability of a random imitation of a peak with such 
amplitude for the temporal resolution corresponding to 5-year observations is $\sim 2\%$ for 27 h 
and $<0.01\%$ for 13.5 h. The lack of output data does not allow to determine the period with 
accuracy better than $\pm 2$ h, but in theses error limits the revealed period 13.3 h coincides with 
one half of the orbital period known from the optical photometry (see panel a in the Fig 8).
To analyse the phase curve which is characterised the 13.3-h periodicity the ASM RXTE 
outputs obtained for the outburst decrease state were cleared from the slow varying trend and then 
were presented on the diagram, on which measured values of 4U1543-47 X-ray emission were 
plotted in dependence of corresponding part of a phase of 27-h periodicity (see panel b in the Fig 
8). We can see the dropped out points on the certain part of the full phase on this panel. The points 
corresponding to the lower intensity are grouped near two approximately equal phase ranges. This 
confirms the reality of 13.5-h periodicity in the hard emission of 4U1543-47 and indicates that only 
13.5 h is the value of the main period in hard emission. The mean light curve of 13.3-h periodicity 
in hard X-rays which was obtained in the Prognoz 9 experiment is also presented in Fig. 8 (see 
panel c).

\section{DISCUSSION}

Most of the periodic sources of hard emission, which were observed in Prognoz 9 and GRIF 
"Mir" experiments, are characterized by non-eclipsing mean light curves. An exception is the well-
known X-ray eclipsing binary 4U1700-37. Nevertheless there is no doubt that most of these periodicities 
have an orbital origin. From this point of view, it is of particular interest that except 4U1700-37 and 
4U1755-33 all identified sources of hour and day periodicity (H1705-22, GRO J1655-40, 4U1543-
47, Cen X-4) appear to belong to transient objects. Because sources of 98-h and 67-h periodicities 
were not observed in the GRIF experiment, we can assume that they were not active in 1995-1997, 
which could be the consequence of strong variability of their hard emission and, hence, the transient 
character also. 
We can also see that among the periodic sources with periods in day range there are three X-
ray novae (Ophiuchis 1977; Scorpius, 1994, 4U1543-47), which are known as black-hole 
candidates. However, there were no direct indications of overall activity of X-ray nova Ophiuchis 
1977 and Scorpius, 1994 as well as 4U1543-47 in 1983-1984. Detection of the above mentioned 
periodicities was made when their activity was not very large. Thus, we can conclude that some X-
ray transients - black hole candidates between outbursts can be active at the level $\sim 100$ mCrab, 
sufficient to detect their hard emission by sensitive detectors. Moreover, some kind of orbital, but 
non-eclipsing periodicity could be observed in high-energy emission in the time of "intermediate" 
activity of these objects, i.e. such X-ray transients could be not quite quiescent between outbursts. It 
is very likely that this periodicity can be caused by features of the accretion disk surrounding the 
black hole and may be inherent to other sources of X-ray transient - black hole candidate.
To the present time there were no clear observations of the development of an orbital 
periodicity in hard emission. Our observations of non-eclipsing orbital periodicity in hard X-rays 
from black-hole candidates just when they were not outburst-like active could be connected with 
some features of the zones of the X-ray formation in these objects. This can be discussed in some 
details using the example of GRO J 1655-40.
The Prognoz 9 and GRIF "Mir" observations allow to conclude that there is a strict 
periodicity in the hard emission from GRO J 1655-40, which is traceable during difference cycles of 
the source's activity separated by a long time interval ($\sim 12$ years). This periodicity is 
apparently attributable to the orbital motion of the binary's components. The energy spectrum of the 
62-h periodicity in the emission from GRO J 1655-40 \citep{kud01} agrees with various 
spectral measurements of the total hard flux from this source. Thus, it is consistent with the model 
of a two-component spectrum typical of compact binaries, black-hole candidates \citep{sha73}. 
The observation of the 62-h periodicity in the hard emission from GRO J 1655-40 
during different epochs suggests that even between its X-ray outbursts, the source was not 
completely quiescent in the hard energy range. In particular, it was active between the X-ray 
outbursts in August 1995 and February 1996. This gives some indications that it could be possible 
to reduce the typical recurrence time, which is important for estimating the mean rate mass transfer 
from the binary's optical companion tip to its compact object $\dot M_T$ \citep{esi00} by at 
least half. The GRIF data also suggest, that taking the source's distance of $\sim 3.5$ kpc \citep{hje95}, 
its luminosity between X-ray outbursts in the periodicity of hard emission at 
energies $E \geq 25$ keV is $\sim 4 \cdot 10^{35}$ erg/s, which is more than three orders of magnitude 
higher than the luminosity that has been believed to be typical of the quiescent state of such objects 
\citep{esi00}.

The intensity of the periodic component of GRO J1655-40 in soft X-rays obtained from 
ASM RXTE observations increases with the increase of the full intensity. It reaches a $3\%$ level 
of the full flux in the whole range of its values. On the other hand, there are some indications on that 
during the time, when the source brightness decreasing after outbursts, in particular, between 
outbursts of 1996 and 1997 years the periodic component in hard X-rays was not low, but on the 
contrary was becoming more contrast. The results of GRIF "Mir" observations indicate that orbital 
periodicity revealed in the GRO J1655-40 hard emission even in the interval between time of its 
high X-ray activity. In this case the emission of GRO J1655-40 in the phase of minimum intensity has 
not fallen to zero. It should be noted that high flux variability during epochs of the source high 
activity makes it difficult to reveal the periodic component with high significance. The source's 
variability is associated not only with periodic variations because the rms deviations of the 
instrument's outputs in all the intervals of observations are several times higher than the amplitude 
of periodic component, which can be revealed only due to averaging over many periods. 
The BATSE CGRO data revealed no manifestations of the periodicity in the hard emission 
from GRO J1655-40 \citep{zha96}. This could be because the amplitude of the 62-h 
periodicity in the hard energy range weakly depends on the total flux. In this case, the periodicity 
would show up most clearly when the source's activity is relatively low (to all appearances, the 
Prognoz 9 and GRIF "Mir" observations occurred precisely at such epochs). However, during 
outbursts, the total flux can increase by more than an order of magnitude. This severely hampers the 
observation of the periodicity, because it proves to be strongly noised by fluctuations in the total 
flux. 
Thus, we can conclude, that although the time profiles of mean light curves, which 
characterize the periodic component in GRO J1655-40 emission are similar in soft and hard X-rays, 
62-h periodicity manifests quite different behaviour in these energy ranges. This could be caused not 
only by different physical mechanisms of generation of soft and hard X-rays in the source, but even 
by emission of hard and soft X-rays by different regions of the accretion disk. This assumption does 
not contradict the spectral features of GRO J1655-40 hard emission, which consists of, at least, two 
or more components. The soft thermal component, which is dominant at energies less than 10 keV 
and hard non-thermal "comptonisation" component, which becomes dominant at higher energies, 
were revealed clearly. It also should be noted that the minimum on the mean phase dependence as in 
soft as in hard X-rays is displaced relatively to the phase zero of the optical light curve. Such non-
coincidence of optical and X-ray light curves for emission periodic components caused by the 
orbital motion could be typical not only for GRO J1655-40. For example, we saw before, the 
periodic component of 4U1543-47 emission also demonstrates different behaviour in optical and X-
ray ranges. The optical light curve gives the orbital period, which is twice larger than the period 
observed in X-rays. The quasi-ellipsoidal form of the star-companion of a binary system can explain 
the non-coincidence of these periods. In any event, the features mentioned above as well as the 
complicated form of mean light curves of periodic component in hard X-rays give no opportunity to 
explain the observed periodicity in X-ray binaries - black-hole candidates using the simple eclipsing 
geometry in orbital motion of binary system companions. It indicates that processes, which provide 
such periodicity, depend also on the accretion disk dynamics, and they may reflect the accretion disk 
structure non-uniformity.
Thus, the main conclusions drawn from observations of periodic processes in hard emission 
of X-ray binaries - black-hole candidates, which were made on Prognoz 9 and GRIF "Mir" 
experiments with the use of data of CGRO and RXTE observations are:
- periodic processes of days range of periods are inherent to the transient source like a X-
ray Nova - black-hole candidate;
- such periodicity is orbital but its origin is not connected with eclipses of the compact 
companion of a binary system;
- periodic processes in soft and hard X-rays are associated with orbital motion of binary 
system companions, but the physical mechanisms and the regions of generation of soft 
and hard X-rays are quite different.

\clearpage
\begin{figure}
\plotone{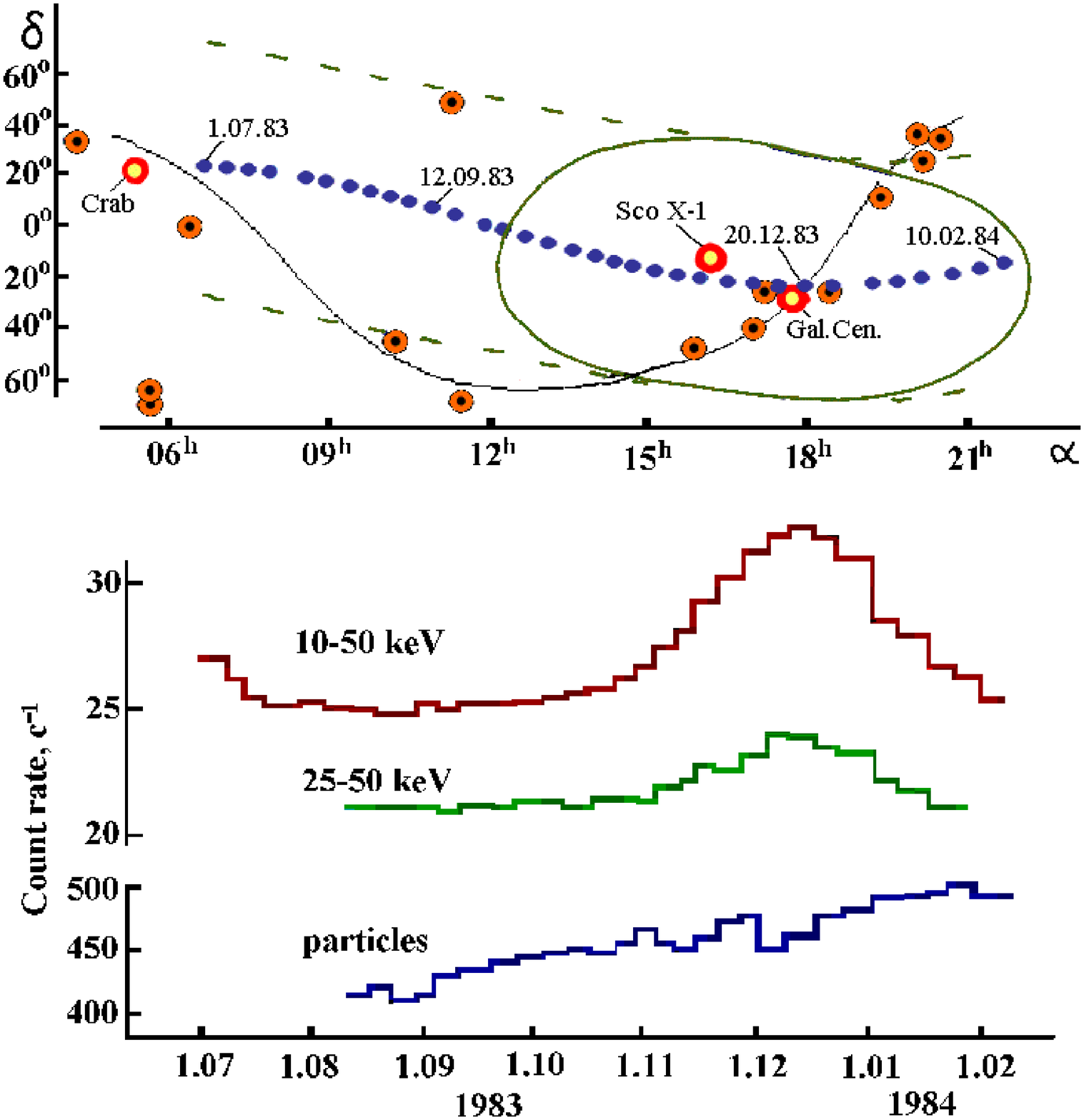}
\caption{a. The part of the sky observed in experiment onboard "Prognoz 9"
satellite. The instrument's field of view, the points of the sky where 
the instrument was directed and some of the bright x-ray sources
and black hole candidates are plotted. b. The mean for the periods of
nonchanging orientation count rates in x-ray 10-25 keV and 25-50 keV 
channels as well as the count rate of anticoincidence shield ("particles").
\label{fig.1}}
\end{figure}
\clearpage
\begin{figure}
\plotone{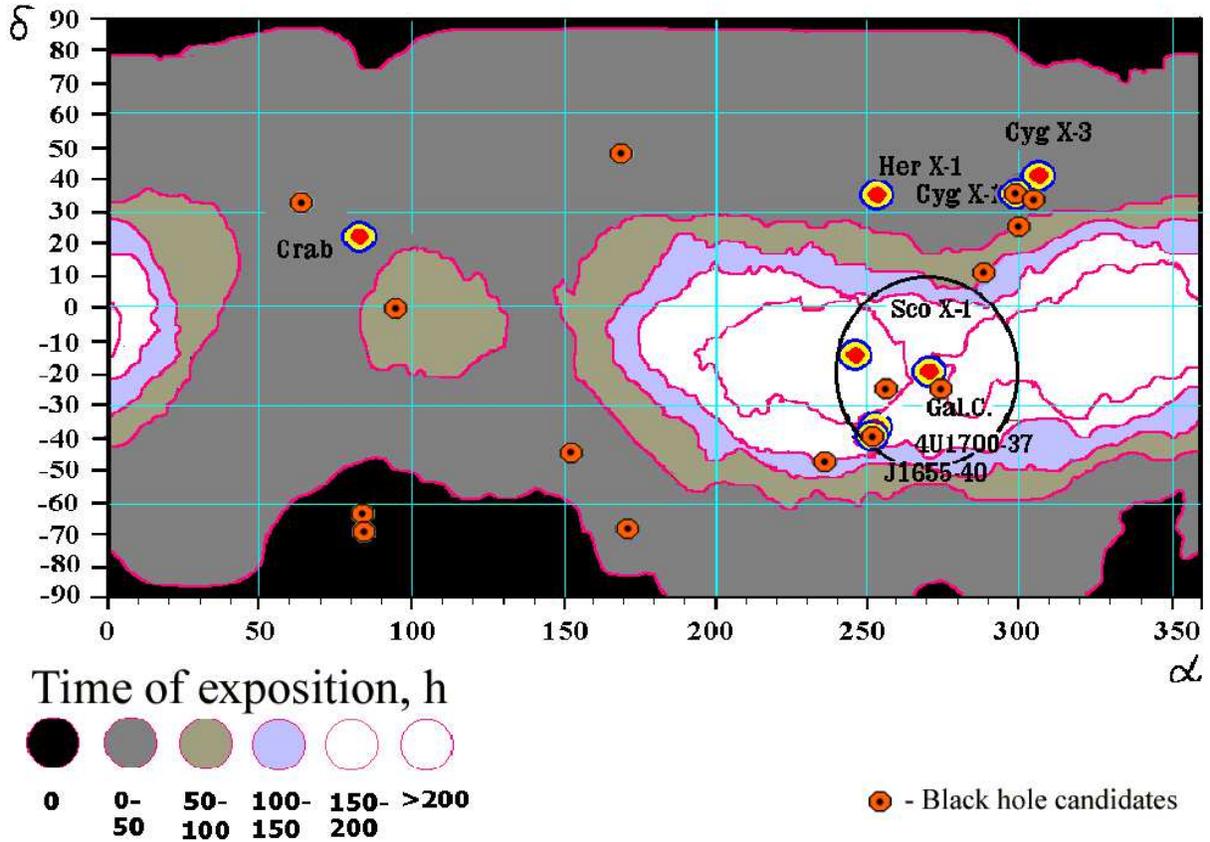}
\caption{
The part of the sky observed with RX-2 instrument in "GRIF" experiment
onboard "Mir" orbital station. Brightness corresponds the time of exposition. 
Bright x-ray sources and black hole candidates are placed.The instrument's 
field of view (on HWHM of its diagram) when it is oriented to the Galactic 
Center is marked as a circle.\label{fig.2}}
\end{figure}
\clearpage
\begin{figure}
\plotone{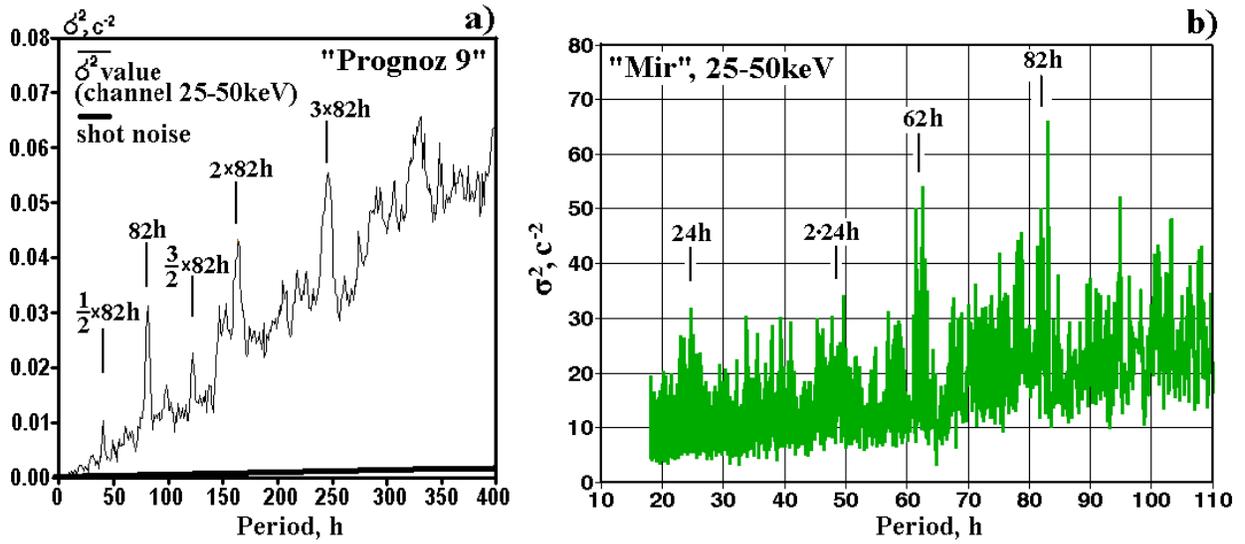}
\caption{a. The periodogram obtained by analyzing the hourly mean
photon count rates in the channel 25-50 keV during the observations
of the Galactic center region (October 31,1983 - January 12,1984).
The peaks in the periodogram correspond to the 82-h period (the orbital
period of the X-ray binary 4U 1700-37) and its multiples.
b. The periodogram was obtained by analyzing the data of "GRIF" experiment
(RX-2 instrument) onboard "Mir" station. The instrument was oriented
to 4U1700-37 and GRO J1655-40 sources with 82-h and 62-h orbital periods 
correspondingly. The wide features in the regions of 24-h multiples are 
connected with day variations of background on "Mir" station orbit. 
\label{fig.3}}
\end{figure}
\clearpage
\begin{figure}
\plotone{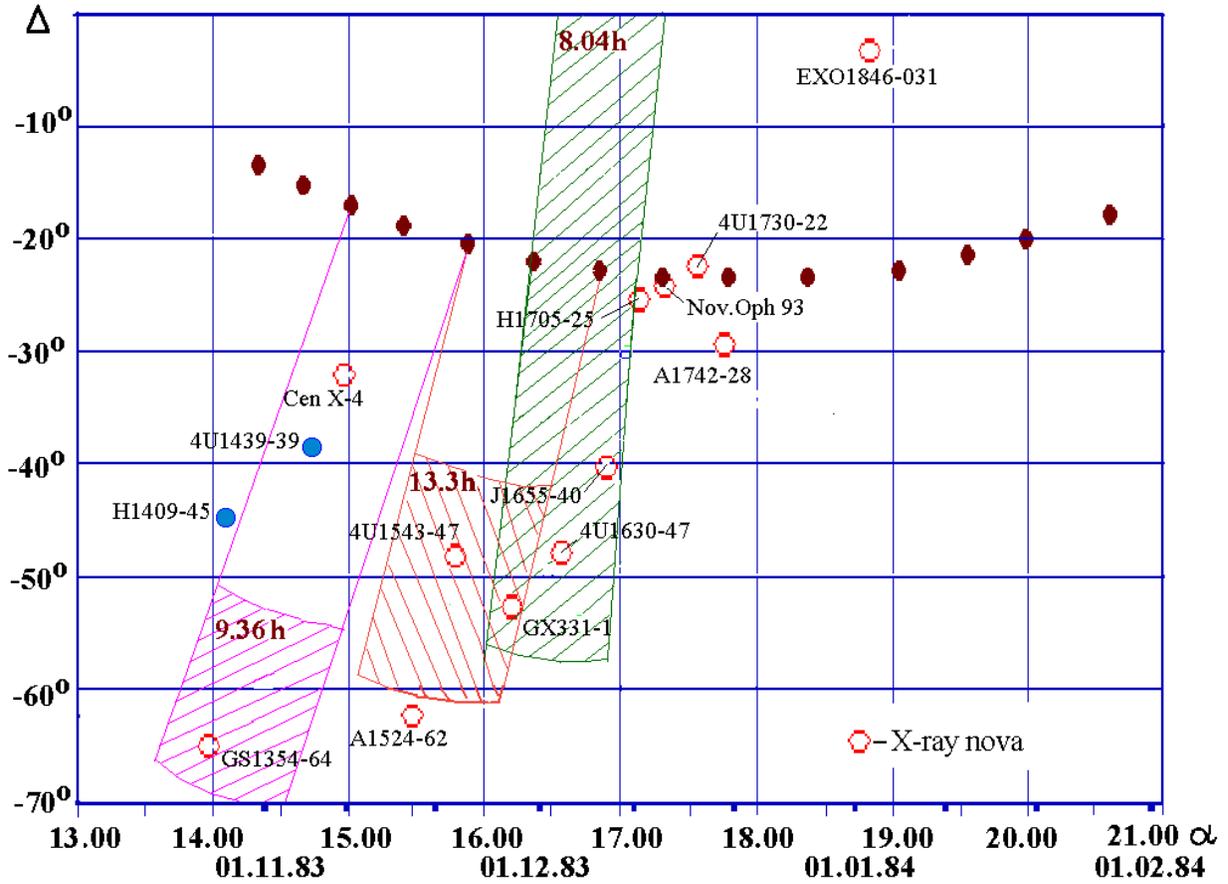}
\caption{The part of the sky in equatorial coordinates with error boxes for 
the sources of 8.04, 9.36, 13.3 h periodicities. The empty circles mark the 
X-ray novaes, the filled circles mark the Prognoz-9 instrument's field center
in the corresponding time.
\label{fig.4}}
\end{figure}
\clearpage
\begin{figure}
\plotone{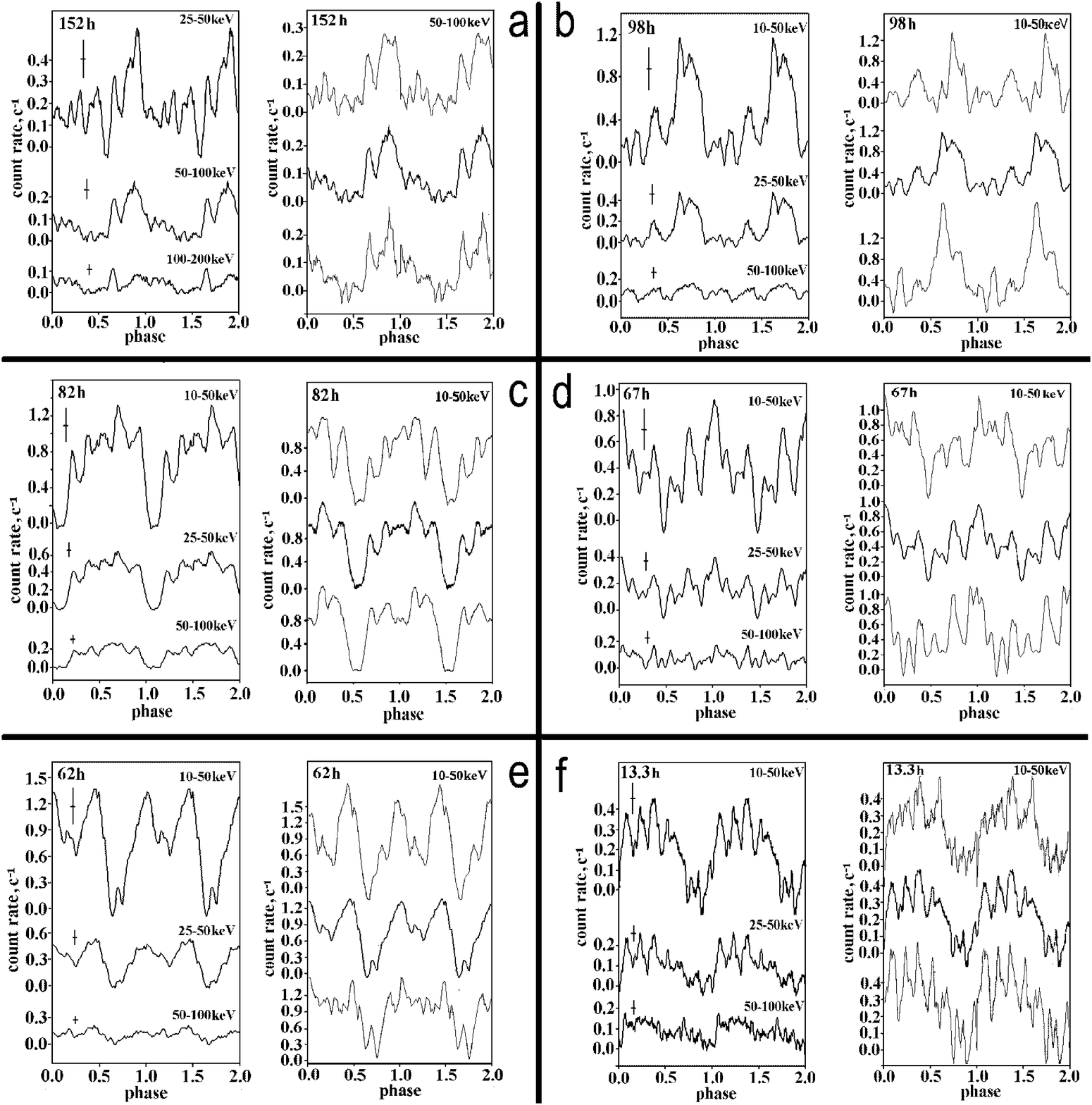}
\caption{The mean light curves corresponding to the periodic sources observed 
during Prognoz-9 mission: 
a. 152 h (time of observation 31.10.1983 - 12.01.1984), 
b. 98 h (08.12.1983 - 06.01.1984), c. 82 h (08.12.1983 - 06.01.1984),
d. 69 h (08.12.1983 - 06.01.1984), e. 62 h (13.11.1983 - 20.12.1983),
f. 13.33 h (08.12.1983 - 06.01.1984). 
The curves are presented for the energy ranges where the corresponding 
periodicities are the most contrast. 
\label{fig.5}}
\end{figure}
\clearpage
\begin{figure}
\plotone{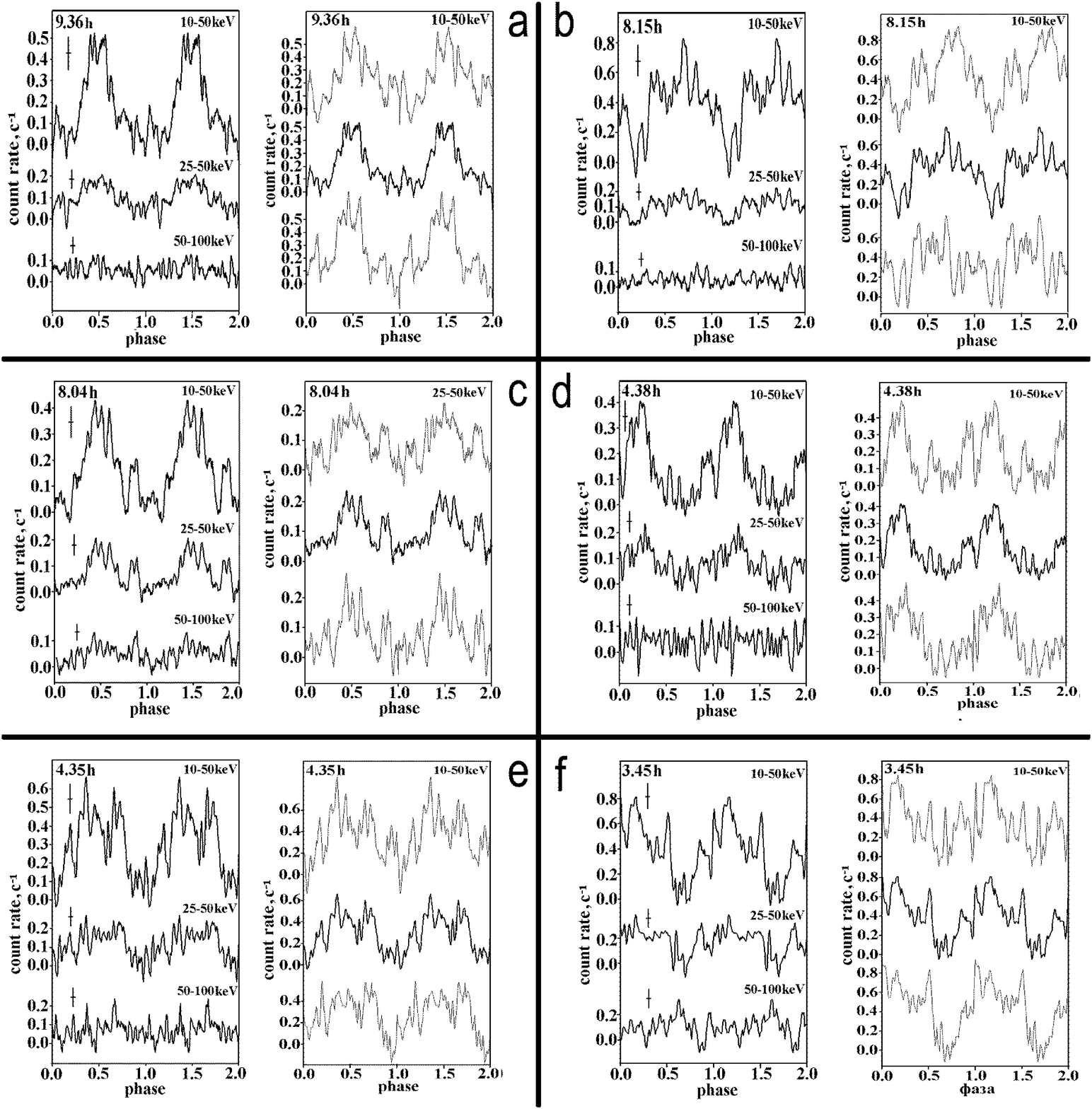}
\caption{The mean light curves corresponding to the periodic sources observed 
during Prognoz-9 mission: 
a. 9.36 h (time of observation 28.10.1983 - 25.11.1983),
b. 8.15 h (13.11.1983 - 25.11.1983), c. 8.04 h (08.12.1983 - 12.01.1984),
d. 4.38 h (28.10.1983 - 10.11.1983), e. 4.35 h (20.12.1983 - 27.12.1983),
f. 3.45 h (07.12.1983 - 13.12.1983).
The curves are presented for the energy ranges where the corresponding 
periodicities are the most contrast. 
\label{fig.6}}
\end{figure}
\clearpage
\begin{figure}
\plotone{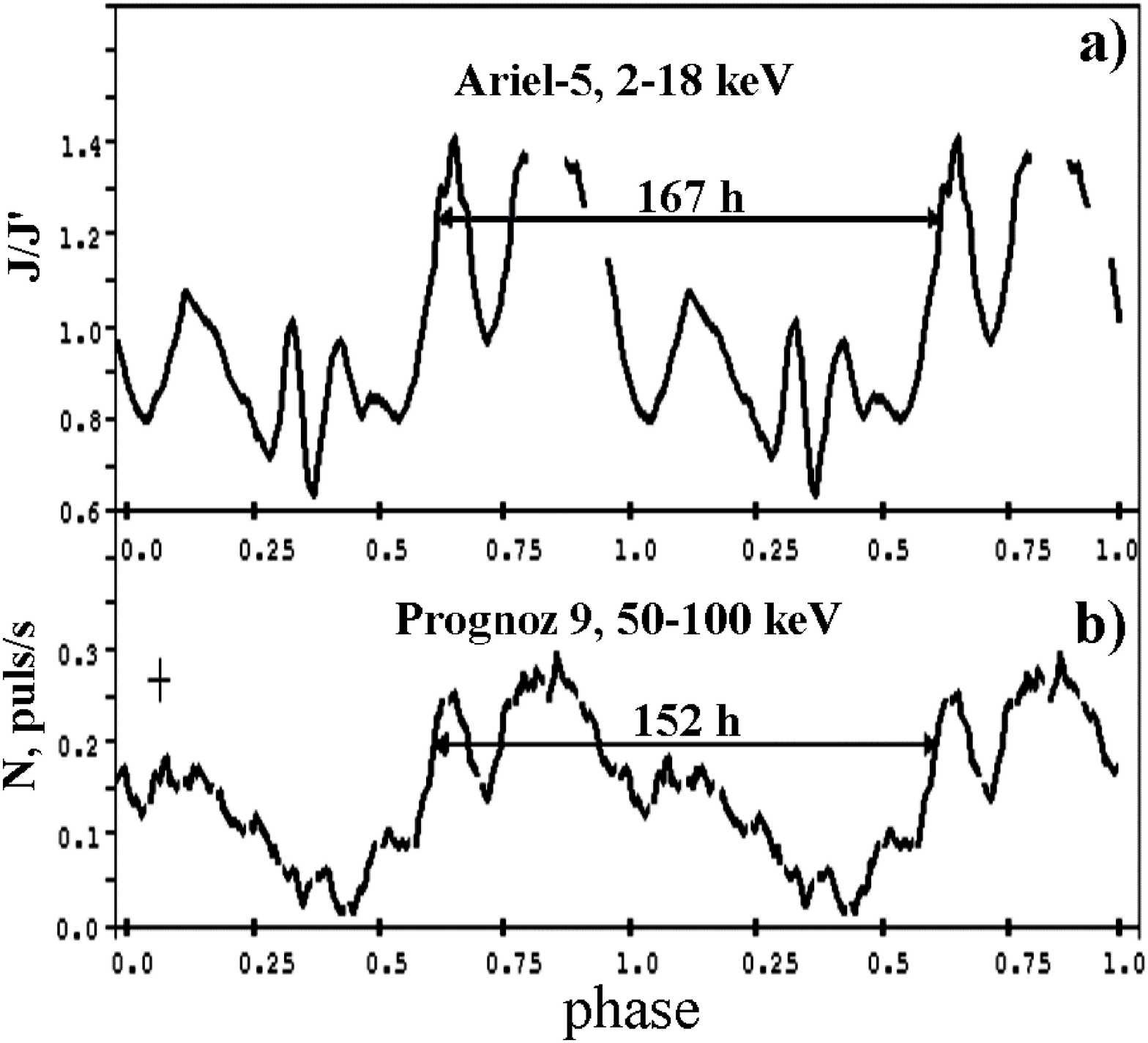}
\caption{The mean phase dependencies obtained on the base of Prognoz 9 
(152 h period) and Ariel 5 SSI \citep{wat78} (167 h period)  
data.\label{fig.7}}
\end{figure}
\clearpage
\begin{figure}
\plotone{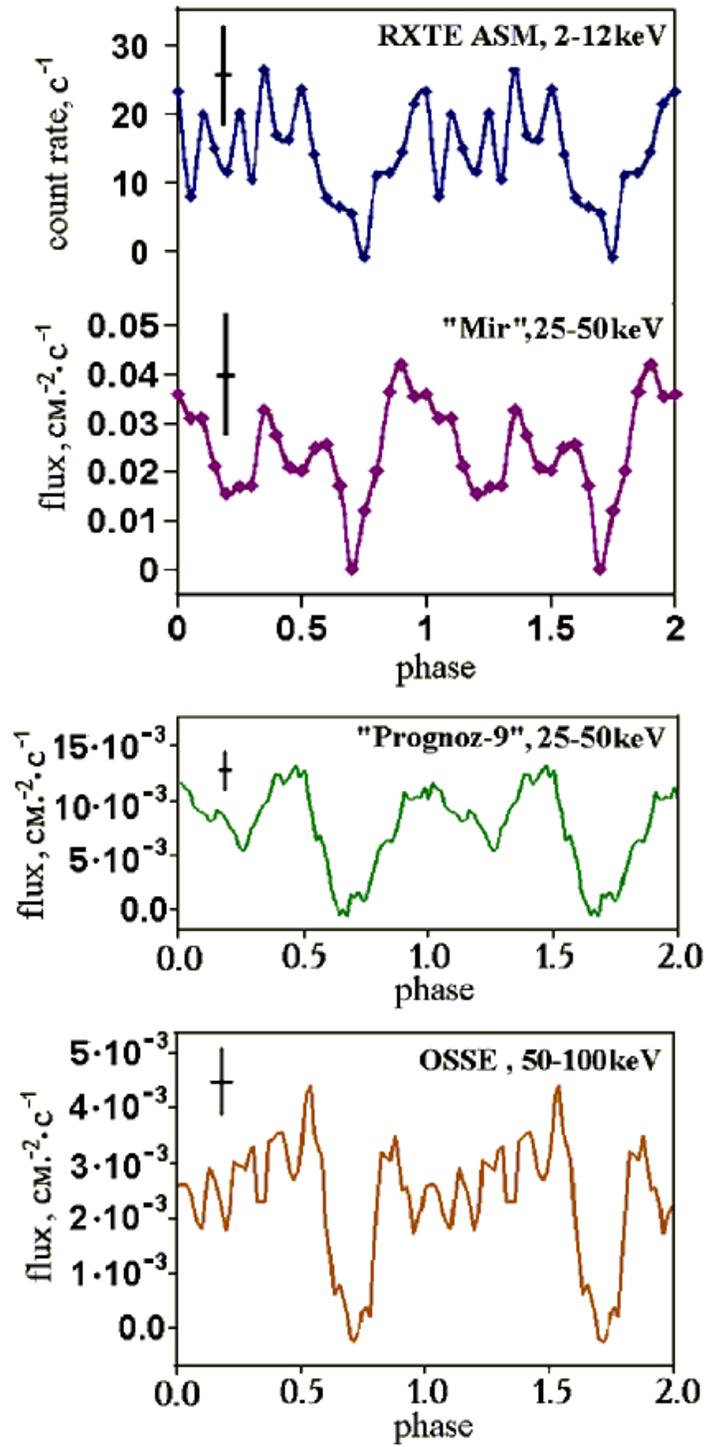}
\caption{The mean light curves for the orbital periodicity of GRO J1655-40
obtained from the ASM RXTE, GRIF OS "Mir", Prognoz 9 and OSSE CGRO 
experiments.\label{fig.8}}
\end{figure}
\clearpage
\begin{figure}
\plotone{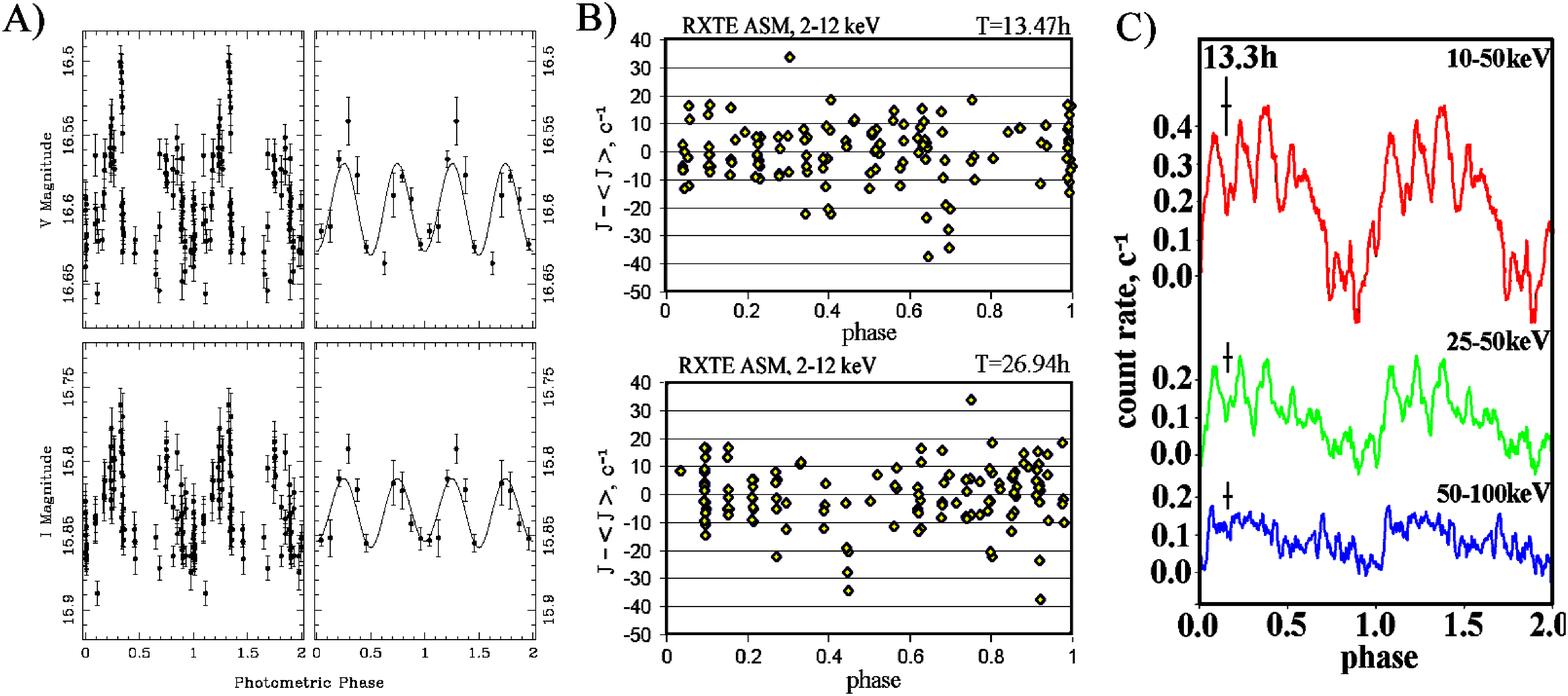}
\caption{a. Optical light curves of 4U1543-47. b. The dependencies of
the intensity of 4U 1543-47 X-radiation measured by ASM RXTE during the
flare in 2002 on 13.47-h and 26.94-h phase (1/2 of orbital period and
orbital one)
\label{fig.9}}
\end{figure}
\clearpage
\begin{deluxetable}{ccrcc}
\tablecaption{Astrophysical sources of periodic X-ray emission observed 
during the Prognoz 9 experiment\label{tbl-1} }
\tablewidth{0pt}
\tablehead{
\colhead{$T$, h} & \colhead{kT , keV} & 
\colhead{$T_1 , T_2$ \tablenotemark{a}} & 
\colhead{N(T) \tablenotemark{b}} &
\colhead{Identification}
}
\startdata
$152 \pm 7$ & $>40$ & 31.10.1983 - 12.01.1984 & 11.5 & H 1705-25 (Nov Oph 1977)\\
$98\pm 7$ & 13 & 08.12.1983 - 06.01.1984 & 7 & \\
$82\pm 5$ & 25 & 08.12.1983 - 06.01.1984 & 8.5 \tablenotemark{c} & 4U 1700-37\\
$69\pm 4$ & 6 & 08.12.1983 - 06.01.1984 & 10 & \\
$62\pm 2$ & 10-20 & 13.11.1983 - 20.12.1983 & 14 & GRO J 1655-40 (Nov Sco 1994)\\
$13.33\pm 0.13$ & 25 & 08.12.1983 - 06.01.1984 & 52 & 4U1543-47\\
$9.36\pm 0.07$ & 10 & 28.10.1983 - 25.11.1983 & 71 & \\
$8.15\pm 0.11$ & 7 & 13.11.1983 - 25.11.1983 & 35 & Cen X-4\\
$8.04\pm 0.04$ & 20-45 & 08.12.1983 - 12.01.1984 & 104 & \\
$4.38\pm 0.03$ & 11 & 28.10.1983 - 10.11.1983 & 71 & \\
$4.35\pm 0.05$ & $<5$ & 20.12.1983 - 27.12.1983 & 33 & 4U 1755-33 \\
$3.45\pm 0.04$ & 5 & 07.12.1983 - 13.12.1983 & 42 & \\
\enddata
\tablenotetext{a}{The interval of best visibility}
\tablenotetext{b}{The number of periods in the interval of best visibility}
\tablenotetext{c}{Total $\approx 20$}
\end{deluxetable}

\clearpage
\begin{deluxetable}{rrrr}
\tablecaption{The upper limits on the intensity of 67, 98, 152 h periodicities as observed in the GRIF experiment. \label{tbl-2} }
\tablewidth{0pt}
\tablehead{
\colhead{Period} & \colhead{25-50 keV} & \colhead{50-100 keV} & \colhead{100-200 keV}
}
\startdata
67 h & $<4.6\cdot10^{-4} cm^{-2}s^{-1}keV^{-1}$ & $<7.2\cdot10^{-4} cm^{-2}s^{-1}keV^{-1}$ & $<2.5\cdot10^{-4} cm^{-2}s^{-1}keV^{-1}$ \\
98 h & $<5.8\cdot10^{-4} cm^{-2}s^{-1}keV^{-1}$ & $<9.0\cdot10^{-5} cm^{-2}s^{-1}keV^{-1}$ & $<3.1\cdot10^{-4} cm^{-2}s^{-1}keV^{-1}$ \\
152 h & $<9.1\cdot10^{-4} cm^{-2}s^{-1}keV^{-1}$ & $<2.6\cdot10^{-4} cm^{-2}s^{-1}keV^{-1}$ & $<2.2\cdot10^{-4} cm^{-2}s^{-1}keV^{-1}$ \\
\enddata
\end{deluxetable}

\end{document}